\documentclass[useAMS,usenatbib]{mn2e}

\usepackage{graphicx}
\usepackage{amssymb}
\usepackage{ulem}

\title[Orbit and masses of $\delta$ Sco]{A new determination of the orbit and
 masses of the Be binary system $\delta$ Scorpii}

\author[W. J. Tango et al.]
 {W. J. Tango,$^1$\thanks{Email: W.Tango@physics.usyd.edu.au}
  J. Davis,$^1$
  A. P. Jacob,$^1$
  A. Mendez,$^1$
  J. R. North,$^1$
  \newauthor
  J. W. O'Byrne,$^1$
  E. B. Seneta$^2$ and
  P. G. Tuthill$^1$\\
  $^1$Sydney Institute for Astronomy, School of Physics, University of Sydney, NSW 2006, Australia\\
  $^2$Cavendish Laboratory, Univ. of Cambridge, UK
}
\date{xxxx xxxx}

\pagerange{\pageref{firstpage}--\pageref{lastpage}} \pubyear{2008}

\begin{document}

\label{firstpage}

\maketitle

\begin{abstract}
The binary star $\delta$ Sco (HD143275) underwent remarkable
brightening in the visible in 2000, and continues to be irregularly
variable. The system was observed with the Sydney University Stellar
Interferometer  (SUSI) in 1999, 2000, 2001, 2006 and 2007. The 1999
observations were consistent with predictions based on the previously
published orbital elements.  The subsequent observations can only be
explained by assuming that an optically bright emission region with an
angular size of $\gtrsim2\pm1$\,mas formed around the primary in 2000.
By 2006/2007 the size of this region grew to an estimated
$\gtrsim4$\,mas.

We have determined a consistent set of orbital elements by
simultaneously fitting all the published interferometric and
spectroscopic data as well as the SUSI data reported here. The
resulting elements and the brightness ratio for the system measured
prior to the outburst in 2000 have been used to estimate the masses of
the components. We find $M_A = 15 \pm 7 M_\odot$ and $M_B = 8.0 \pm 3.6
M_\odot$. The dynamical parallax is estimated to be $7.03\pm0.15$\,mas,
which is in good agreement with the revised HIPPARCOS parallax.\end{abstract}

\begin{keywords}
 binaries: spectroscopic --- binaries: visual --- stars: fundamental parameters
--- stars: individual ($\delta$ Sco, HR 5953) --- techniques: interferometric
\end{keywords}

\section{Introduction}\label{sec:intro}

The bright southern star $\delta$ Sco (HR 5953, HD143275; RA $=16^{\rm
h}~00^{\rm m}~20.\!\!^{\rm s}01$, $\delta=-22^\circ\ 37^\prime\
18^{\prime\prime}$) is listed in the Bright Star Catalogue (BSC)
\citep{hoffleit91} as a B0.3\,IV star.  It is listed in the Sixth Catalog of
Visual Binary Stars \citep{hartkopf06} as a spectroscopic triple and
occultation quadruple system; however, the overwhelming evidence is that it is
a binary system with very high eccentricity.

In 2000 \citet*{otero01} observed a remarkable brightening of this star
and since then it has exhibited irregular variability. It is now
classed in SIMBAD as a B0.2\,IVe star, and it has been suggested that
it may be a $\gamma$ Cas type variable \citep{otero01}.

We report here observations made with the Sydney University Stellar
Interferometer (SUSI) from 1999 to 2007 that are consistent with the
development of an optically thick circumstellar disk around the primary star
and discuss the implications for our understanding of $\gamma$ Cas variable
stars.

The orbital elements for $\delta$ Sco have been determined both from
interferometry and spectroscopy (see Section \ref{sec:history}) but the
elements found by the the two techniques are inconsistent. Following
\citet{pourbaix98} we present a new analysis of the orbital and radial
velocity (RV) data that simultaneously minimises the residuals in $x$,
$y$ and $\dot{z}$, where $x$, $y$ are the cartesian coordinates of the
secondary with respect to the primary projected onto the plane of the
sky and $\dot{z}$ is the heliocentric radial velocity of the secondary.
This new solution is consistent with both the interferometric and
spectroscopic data and allows us to estimate the masses of the A and B
components of $\delta$ Sco as well as the dynamical parallax of the
system.

\section{Observations prior to 1997}\label{sec:history}

\subsection{Lunar occultation and interferometric observations}

The binary nature of $\delta$ Sco was first  reported by
\citet{innes1901}, who observed it  during a lunar occultation in 1899. This
work was largely forgotten until 1974 when $\delta$ Sco was
rediscovered to be binary using lunar occultation \citep{dunham74}, by
interferometry with the Narrabri Stellar Intensity Interferometer
(NSII) \citep*{hbrown74} and by speckle interferometry
\citep{labeyrie74}. Since then it has been regularly observed with
speckle interferometry \citep[see][for references]{hartkopf04}.

\citet{bedding93} observed $\delta$ Sco with the Masked Aperture-Plane
Interference Telescope (MAPPIT). Using his measurement and the measures
previously obtained with speckle interferometry he was able to calculate an
orbit for the system.  The MAPPIT data were also used to estimate the
brightness ratio   of the two components.

The orbital elements were recalculated
by \citet*{hartkopf96}.%
 \footnote{This work noted that the measurement of \citet{bedding93} ``proved discrepant''
 and was not included in the orbital fit.  This was based on a misreading of the
 epoch of Bedding's observation; the error has been corrected in the Fourth
 Catalog of Interferometric Measurements of Binary Stars \citep{hartkopf04}.}
A revised orbit using more recent spectroscopic data has been
calculated by \citet{miro01} but it is clearly inconsistent with the
interferometric data.

The accurate determination of the orbital elements using interferometry has
proved to be rather difficult. Part of the reason is that $\beta$, the
brightness ratio of the secondary to the primary, is approximately $0.2$. For
interferometric observations the signal-to-noise ratio (SNR)
 for the fringe modulation due to the binary (see equation
(\ref{eq:doublemono})) is  multiplied by a factor of $2\beta/(1+\beta)$ and for
$\delta$ Sco this is approximately 0.3.

The other difficulty is the large eccentricity of the orbit ($e
> 0.9$). It is apparent from Fig.~\ref{fig:orbit} that there are no speckle
observations near periastron since the two stars are too close to be resolved
when using apertures of the order of 5 metres. The original motivation for this
work was to observe $\delta$ Sco near periastron with SUSI using a baseline of
40\,m (corresponding to an angular resolution of $\sim2$\,mas for an observing
wavelength of $442$\,nm).

\subsection{Spectroscopy and photometry}

The BSC \citep{hoffleit91} noted that $\delta$ Sco is a spectroscopic
binary with a period of approximately 20 days
\citep*[see][]{vanHoof81}, although \citet{levato87} found a period
closer to 80 days.  More recent spectroscopy (see below) shows no clear
evidence of a short period companion.

Prior to 1993 $\delta$ Sco had been regarded as a typical B star and
was listed as a photometric standard in several catalogues \citep*{slettebak82,
slettebak85, perry87}. However, \citet{cote93} reported that the H$\alpha$ profile 
showed emission on the flanks of the
absorption line and  they proposed that it should be classified
as a Be star. \citet{bedding93} noted that this observation was made
approximately 10 months after periastron and speculated that the two
events might be related.

\section{The SUSI results}

\begin{figure*}
\begin{center}
\includegraphics[scale=0.6, angle=270]{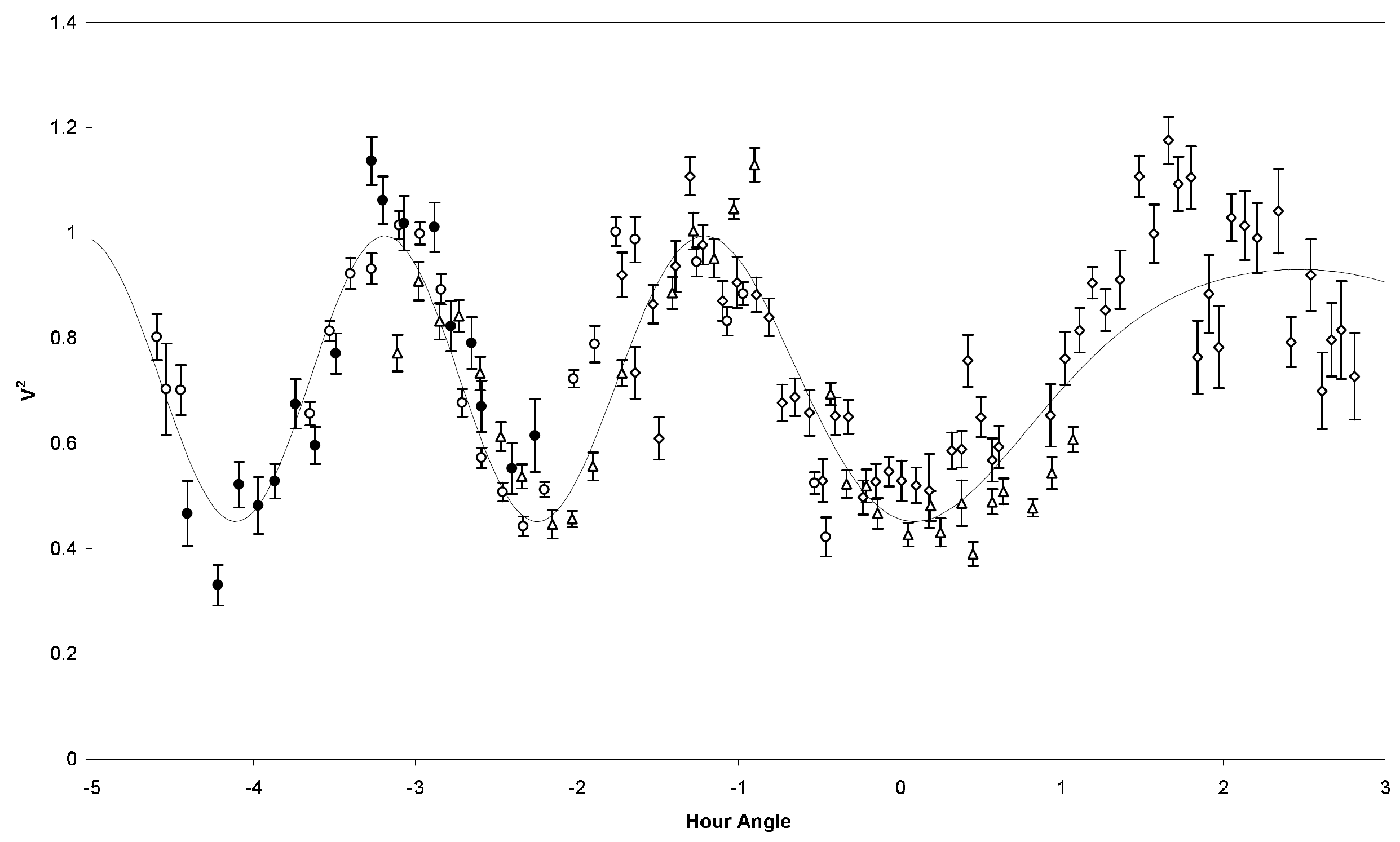}
\caption{The SUSI data for the 1999 observations. The data from the different
nights are shown as follows: 1999 May 29 ($\circ$), 1999 May 30 ($\bullet$),
1999 May 31 ($\diamond$) and 1999 June 06 ($\bigtriangleup$). The line is the
best fit to the weighted data.}\label{fig:merged_data}
\end{center}
\end{figure*}

\subsection{Observations in 1999}

$\delta$ Sco was initially observed with SUSI using the original
``blue'' configuration described in \citet{davis99a}. Like most
two-aperture interferometers, SUSI measures the squared modulus of the
fringe visibility, and for binary stars this is given by
\begin{equation}\label{eq:doublemono}
    |V|^2 = \frac{|V_1|^2 + \beta^2|V_2|^2 -
    2\beta|V_1|\cdot|V_2|\cos\psi}{(1+\beta)^2}
\end{equation}
where
\begin{equation}\label{eq:psi}
\psi =
\frac{2\pi{\mbox{\mathversion{bold}$b$}}\cdot{\mbox{\boldmath$\rho$}}}{\lambda}
\end{equation}
Here $V_1$ and $V_2$ are the visibilities of the primary and secondary
respectively, $\beta \leq 1$ is the brightness ratio,
{\mathversion{bold}$b$} is the baseline vector projected on the sky,
{\boldmath$\rho$} is the vector separation of the two stars, and
$\lambda$ is the wavelength \citep{hbrown70}. As explained elsewhere
\citep{davis05}, due to the difficulty of calibrating binary star
observations with the original, blue-sensitive SUSI instrumentation,
the observed squared visibilities, $|V_{\rm obs}|^2$, were fitted using
\begin{equation}
  |V_{\rm obs}|^2 = (C - C_bt)|V|^2
\end{equation}
where $C$ and $C_b$ are fitting parameters, and $t$ is the time. The parameters
$C$ and $C_b$ reflect the fact that seeing diminishes the fringe visibility and
that the seeing loss in general increases in a roughly linear fashion during
the night.

In the case of $\delta$ Sco it was assumed that the angular diameters of the
individual stars were much smaller than the angular separation and that the
orbital motion during any given night was negligible.  It follows that $|V_1|^2
= |V_2|^2 = 1$ and there are five fitting parameters: the angular separation
$\rho$, the position angle $\theta$, the brightness ratio $\beta$ and the two
empirical fitting parameters $C$ and $C_b$.

$\delta$ Sco was observed on the nights of 29, 30, 31 May and 3 June,
1999 at a wavelength of 442\,nm and a baseline of 5\,m. Based on the
published orbital elements the rates of change in the position angle
and separation were expected to be $~0.\!\!^\circ02$ day$^{-1}$ and
$0.14$\, mas$\cdot$day$^{-1}$, respectively, and the variation over
several nights should have been negligible compared to the
observational uncertainties. The actual data, however, showed
considerable night-to-night variation. It was eventually realised that
this was the result of variable seeing. For each night the data were
fitted using the procedure outlined above. The data were then scaled by
$(C + C_b)^{-1}$ to produce ``unbiased'' values of $|V|^2$, and the
data for the different nights were then plotted together. This
immediately highlighted portions of the data that were badly affected
by seeing. These data points were removed and the data refitted. After
several iterations a consistent set of data was obtained for the four
nights, as shown in Fig. \ref{fig:merged_data}. The best fit values for
the angular separation, position angle and brightness ratio for the
epoch J1999.4 are:
\begin{eqnarray}
\rho &=& 81.13\pm0.15\,{\rm mas}\nonumber\\
\theta &=& 16.\!\!^\circ1\pm0.\!\!^\circ2\nonumber\\
\beta &=& 0.195\pm0.005\nonumber
\end{eqnarray}
The brightness ratio expressed as a magnitude difference is also listed in
Table \ref{table:magdiff} along with the values obtained with the NSII and
MAPPIT.

\begin{table}
\begin{center}
\caption{The magnitude difference between the two components of $\delta$
Sco}\label{table:magdiff}
\begin{tabular}{lclc}
\hline \multicolumn{1}{c}{Epoch} &  \multicolumn{1}{c}{Difference} &  \multicolumn{1}{c}{$\lambda$ (nm)} & Instrument \\
\hline
 $1971$ & $1.9\pm0.4$ & $443$ & NSII\\
 $1991$ & $1.5\pm0.3$ & $600^a$ & MAPPIT\\
 $1999$ & $1.78\pm0.03$ & $442$ & SUSI\\
\hline
\end{tabular}
\medskip\\
$^a$\citet*{marson92}.\\
\end{center}
\end{table}

\subsection{Observations in 2000 and 2001}

Attempts were made in 2000 to observe $\delta$ Sco using a 40\,m
baseline, since the angular separation was expected to be
$\sim\!9$\,mas. There were two nights of data in March, five nights in
May and three nights in June. The characteristic modulation expected
for a binary system (see Fig. \ref{fig:merged_data}) was not seen,
although in the case of the March data the SNR was poor due to bad
seeing.

A similar attempt was made in 2001 (four nights in June and five in
July). The separation was expected to be similar to that observed in
1999, but again the results were negative.

\subsection{Observations in 2006 and 2007}

Since the observations in 2001 SUSI has undergone a major upgrade
\citep{davis07}. In particular, a new  ``red'' fringe detection system
has been developed that uses a scanning mirror to sweep through a range
of optical path. A wide optical bandwidth is used, and in the case of a
single star a ``fringe packet'' is observed centred on the white-light
fringe position. Fourier methods are then used to estimate $|V|^2$. For
the observations reported here a bandwidth of 80\,nm centred on 700\,nm
was used.

\begin{figure}
\begin{center}
\includegraphics[scale=0.3]{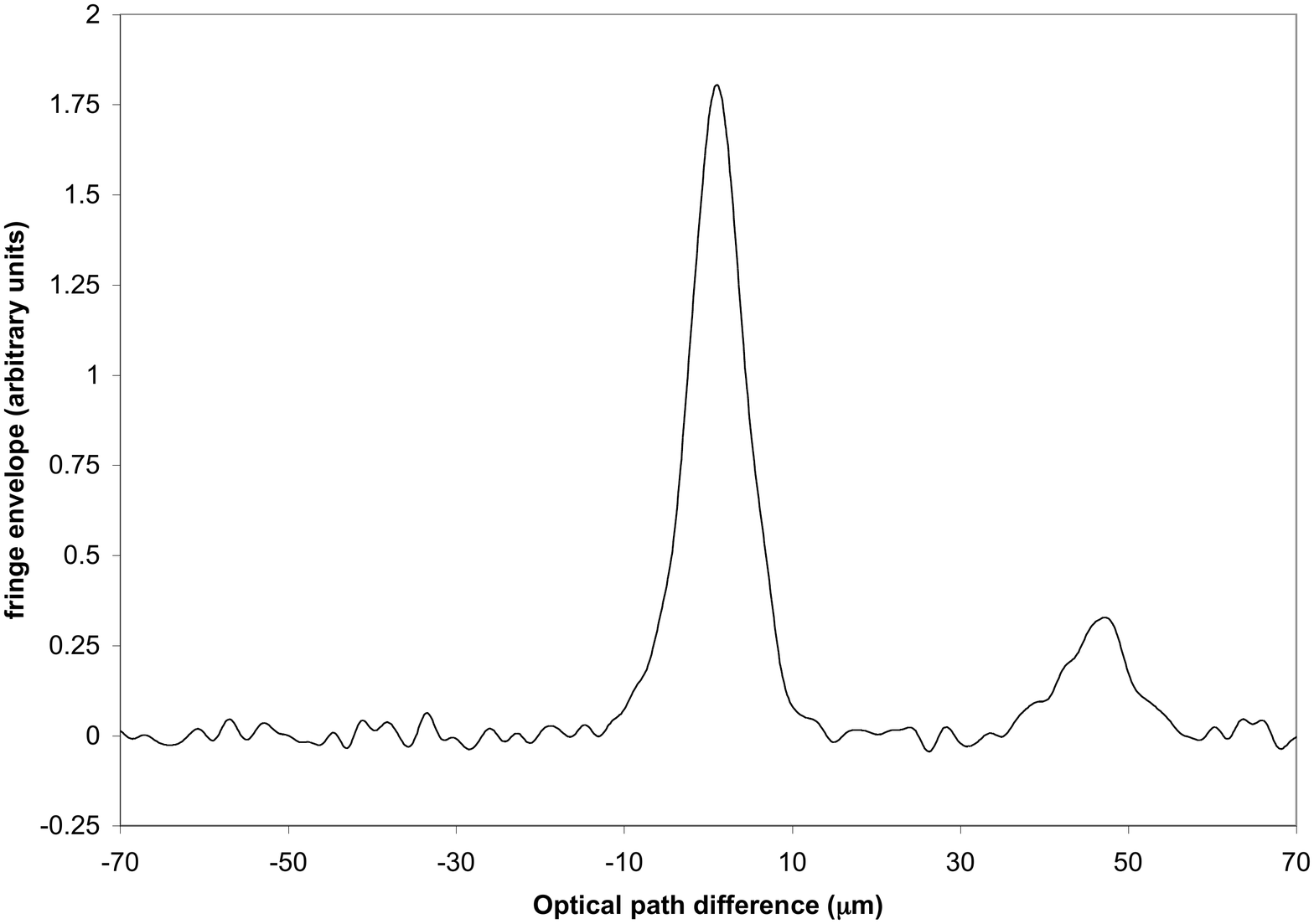}
\caption{The mean fringe envelope for a set of fringe scans taken while
observing the binary star $\alpha$ Dor (HD29305) on 2005 Jan 12 with a 40\,m
baseline at 700\,nm. The envelope due to the primary is located at zero
optical path difference  while that of the secondary is located near $+45\,\mu$m.}\label{fig:alpha_Dor}
\end{center}
\end{figure}

In the case of a binary system, each star will produce its own fringe packet
and the separation between the packets will be $\Delta x = ${\mathversion{bold}
$b\cdot{\mbox{\boldmath$\rho$}}$}.  The width of each fringe packet is
approximately $w = \lambda^2/\Delta\lambda$, where $\Delta\lambda$ is the
bandwidth, and if $\Delta x \ll w$ the two fringe packets will ``beat'' with
each other and the overall fringe visibility will be given by
equation~(\ref{eq:doublemono}).  However, if $\Delta x \gg w$ the two fringe
packets will be separated in optical path difference. Fig.~\ref{fig:alpha_Dor}
shows an example of such a double-peaked fringe
envelope for the binary $\alpha$ Dor.%
\footnote{ The ``fringe envelope'' is the modulus of the analytic signal
associated with the fringe packet. Operationally it can be found using the
expression given by \citet{bracewell99}.  The fringe envelope does not depend
on the phase of the fringe packet and can be averaged over many fringe scans to
improve the SNR.}
 The separation between the envelopes
due to the primary and secondary was approximately $45\,\mu$m when this observation
was made.

In general {\mathversion{bold}$b\cdot{\mbox{\boldmath$\rho$}}$} varies
throughout a set of observations due to the Earth's diurnal rotation.
However, in 2006-2007 the position angle of $\delta$ Sco was
$\sim\!180^\circ$, making it almost parallel with the N-S baseline
configuration of SUSI. As a consequence the separation of the fringe
packets was expected to remain nearly constant throughout an observing
session, making it easy to locate and measure the two fringe peaks.  An
attempt was made to observe the double-peaked signal from $\delta$ Sco
on the night of 30 June 2006 using a 40\,m baseline.  The separation
was estimated to be $\sim\!40\,\mu$m. Only one fringe peak could be
detected.  The absence of the second peak could not be explained by
instrumental effects or the analysis procedure, as the instrumental
configuration was essentially identical to that used for the $\alpha$
Dor observations  and  the position of the secondary peak with
respect to the primary was essentially the same for both stars.
Atmospheric seeing was ruled out because other stars were
observed the same night without any difficulties.

\begin{figure*}
\begin{center}
\includegraphics[scale=0.75]{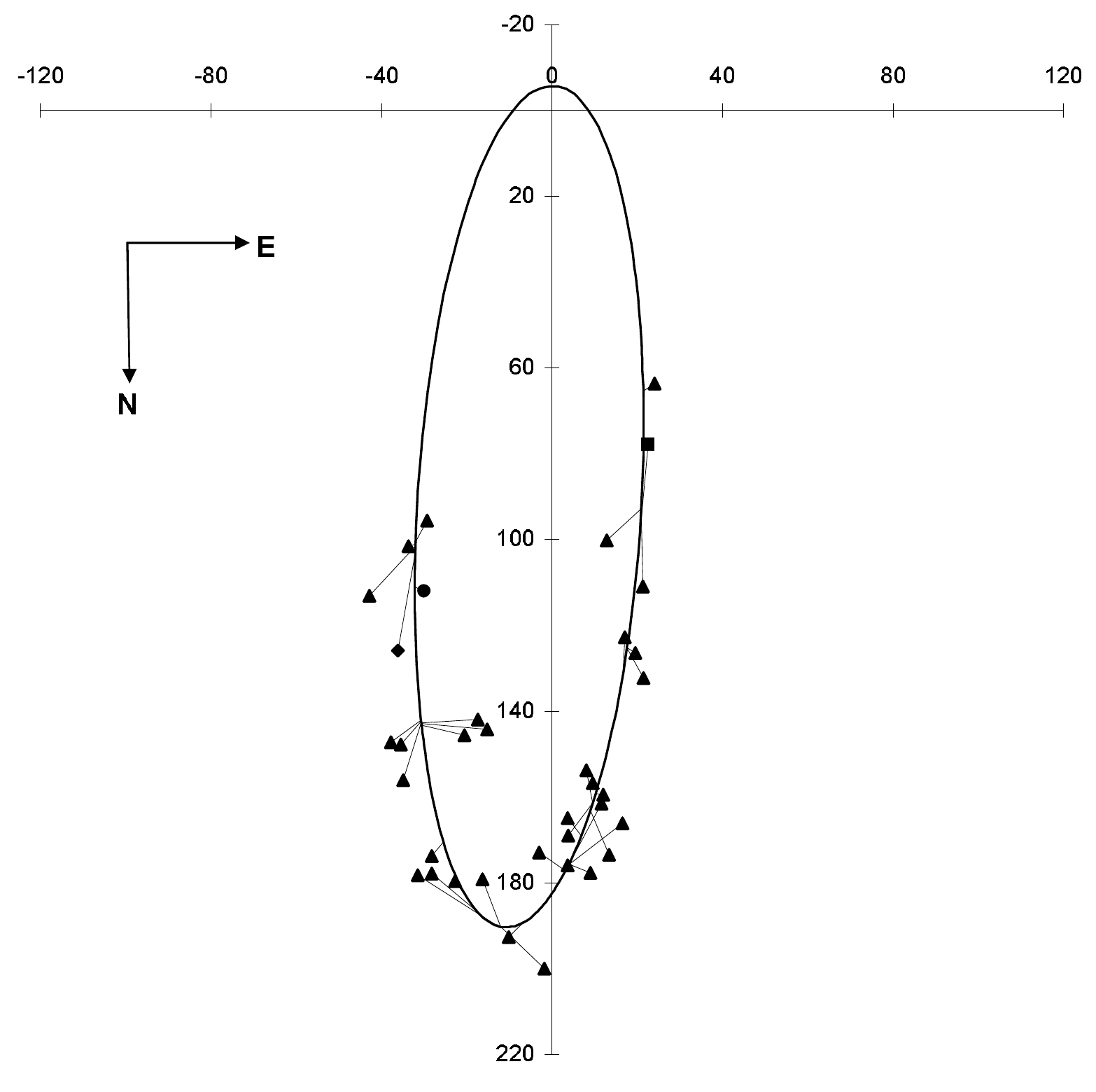}
\caption{The interferometric orbit for $\delta$ Sco using our revised
elements (see Table \protect\ref{table:elements}). The units for both
axes are in mas. The lines indicate the O-C vectors and the symbols
indicate the technique used: $\blacktriangle$ -- speckle
interferometry; $\blacklozenge$ -- HIPPARCOS; $\bullet$ -- MAPPIT;
$\blacksquare$ -- long-baseline optical interferometry (this work).
Except for this latter point all the data are from
\protect\citet{hartkopf04}.\label{fig:orbit}}
\end{center}
\end{figure*}

\subsection{Discussion}

\citet{miro01} explained the spectroscopic and photometric behaviour of
$\delta$ Sco in terms of the development of a circumstellar disk that started to form shortly before
periastron. This scenario explains the brightness
variations that were observed in terms of the complex evolution of this
disk.  This disk may
have been a new phenomenon or may have grown
out of the H$\alpha$ emitting disk seen earlier \citep{miro01,galaz06}.

 The interpretation of the 2005--2006 observations is straightforward.
The fact that only one fringe envelope was seen implies that the
primary was fully resolved; i.e., the angular size of the disk,
projected onto the SUSI baseline, is $\theta_{\rm disk} \gtrsim
1.22\lambda/b = 4.4$\,mas, where $b = 40\,$m is the projected baseline
and $\lambda = 700$\,nm is the wavelength.

The most likely explanation for the apparent negative results obtained
with SUSI in 2000 and later epochs is that the optically thick disk was
fully resolved by the interferometer. This allows us to place lower
limits on the angular size of the disk at the epoch of the
observations.

The 2000--2001 observations were made with the blue beam-combining
system. The effect of partial resolution of the primary would be to
reduce the modulation seen in Fig. \ref{fig:merged_data} even further.
Given the noise in the data we estimate that the modulation would be
undetectable if the primary were $\gtrsim 2\pm1$\,mas.  The relatively
large uncertainty reflects the fact that the actual threshold for
detecting the modulation is not well defined, as it depends on the
seeing conditions.

\citet{carciofi06} have developed a thick disk model to account for the
continuum emission. They estimated the size of the circumstellar disk
in 2001 to be $7R_\star$, where $R_\star$ is the radius of the primary.
Using the data given by \citet{miro01} $R_\star \sim 6 R_\odot$.
Combining this with our dynamical parallax (see Section
\ref{sec:masses}) the size of the disk given by \citet{carciofi06}
corresponds to $2.8$\,mas. This is consistent with our measured value
of $2\pm1$\,mas for the same epoch (2001).

\citet{carciofi06} note that if the disk is geometrically thin and lies
in the orbital plane it will obscure only $\sim\!10\%$ of the stellar
flux, but to explain the observed fluctuations in the light curve one
has to assume that the geometry of the disk must change. If only $\sim\!10\%$ of
the primary is obscured one would expect to see an unresolved component
in the interferometric data taken with the red beam-combining system  in 2005-2006.
The fact that no unresolved component was seen suggests that the disk was in fact
the dominant contribution to the light from the primary plus disk system.

\section{The orbit of $\delta$ Sco}\label{sec:orbit}

\begin{table*}
\begin{center}
\caption{The orbital elements for $\delta$ Sco}.\label{table:elements}
\begin{tabular}{lllll}
\hline \multicolumn{1}{c}{Element} &  \multicolumn{1}{c}{Ref.~{\it a}}&
       \multicolumn{1}{c}{Ref.~{\it b}} &\multicolumn{1}{c}{This Work} \\
\hline
 Period  $P$ (yr) &$10.58\pm0.08$&$10.58^c$&$10.74\pm0.02$\\
 Epoch of periastron  $T$ &B$1971.41\pm0.14$&J$2000.693\pm0.008$& J$2000.69389\pm0.00007$  \\
 Eccentricity  $e$ &$0.92\pm0.02$&$0.94\pm0.01$& $0.9401\pm0.0002$& \\
 Semimajor axis (mas) $a^{\prime\prime}$ &$107\pm7$&$107^c$& $98.3\pm1.2$\\
 Inclination  $i$ &$48.\!\!^\circ5\pm6.\!\!^\circ6$&$38^\circ\pm5^\circ$& $38^\circ\pm6^\circ$ \\
 Long.\ periastron  $\omega$ &$24^\circ\pm13^\circ$&$-1^\circ\pm5^\circ$& $1.\!\!^\circ9\pm0.\!\!^\circ1$ \\
 Long.\ of asc.\ node $\Omega$&$159.\!\!^\circ3\pm7.\!\!^\circ6$&$175^\circ$& $175.{\!\!^\circ}2\pm0.\!\!^\circ6$ \\
 Systemic RV  $V_\gamma$ (km$\cdot$s$^{-1}$)&&$-6\pm0.5$& $-6.72\pm0.05$  \\
 RV amplitude  $K_A$ (km$\cdot$s$^{-1})$ &&& $23.84\pm0.05$  \\
 \multicolumn{2}{l}{Semimajor axis of primary $a_A$ (km)}&& $(7.1\pm0.1)\times10^{8}$  \\
 \multicolumn{2}{l}{Mass function $M_B^3/(M_A+M_B)^2$ (M$_\odot$)} && $0.9\pm0.4$  \\
  \hline
\end{tabular}
\medskip\\
\parbox{0.4\textwidth}{
 Ref.~{\it a} \citet{hartkopf96}\\
 Ref.~{\it b} \citet{miro01}\\
 $^c$ Value assumed from Ref.~{\it a}}
\end{center}
\end{table*}

It is clear from Fig.~5 in \citet{miro01} that their orbit does not fit the
interferometric data as well as the \citet{hartkopf96} orbit. Conversely, their
Fig.~6 shows that the Hartkopf elements do not reproduce their RV curve. It is
apparent from this figure that the spectroscopic data provide a very precise
value for the epoch of periastron $T$. The symmetry of the RV curve around $T$
also implies $\omega$ is very close to zero.

Ideally one should fit the two sets of data simultaneously  and we have
followed the procedure outlined by \citet{pourbaix98} with a few modifications.
Because $\delta$ Sco is a single-lined spectroscopic binary it is natural to
choose $a_A$, the physical semimajor axis of the primary's orbit in kilometres,
as an additional orbital element. For convenience the fitting algorithm
actually optimizes the RV amplitude $K_1$; the semimajor axis is then found
using the standard relation \citep{heintz78}
\begin{equation}\label{eq:a1}
    a_A = \frac{43200f_dPK_A(1-e^2)^{1/2}}{\pi\sin i}
\end{equation}
where $f_d$ is a numerical factor depending on the units of $P$.%
 \footnote{If $P$ is in days, $f_d = 1$. It is equal to $365.25$ or $365.242\ldots$
 if $P$ is in Julian or Besselian years, respectively.}
Similarly we can also derive the mass function in solar masses:
\begin{equation}\label{eq:fm}
    f(m) = \frac{M_B^3}{(M_A+M_B)^2} = 3.985\times10^{-20}\frac{a_A^3}{(f_dP)^2}
\end{equation}
or
\begin{equation}
    f(m)  =  1.03617\times10^{-7}\frac{f_dPK_A^3(1-e^2)^{3/2}}{\sin^3i}
\end{equation}

The \citet{nelder65} simplex algorithm was used to fit the objective function,
which was essentially the same as the one used by \citet{pourbaix98} except
that the last term (the sum of the RV residuals for the secondary) was omitted.
The uncertainties in the elements were estimated by Monte Carlo simulation.

\begin{figure*}
\begin{center}
\includegraphics[scale=0.5, angle = 270]{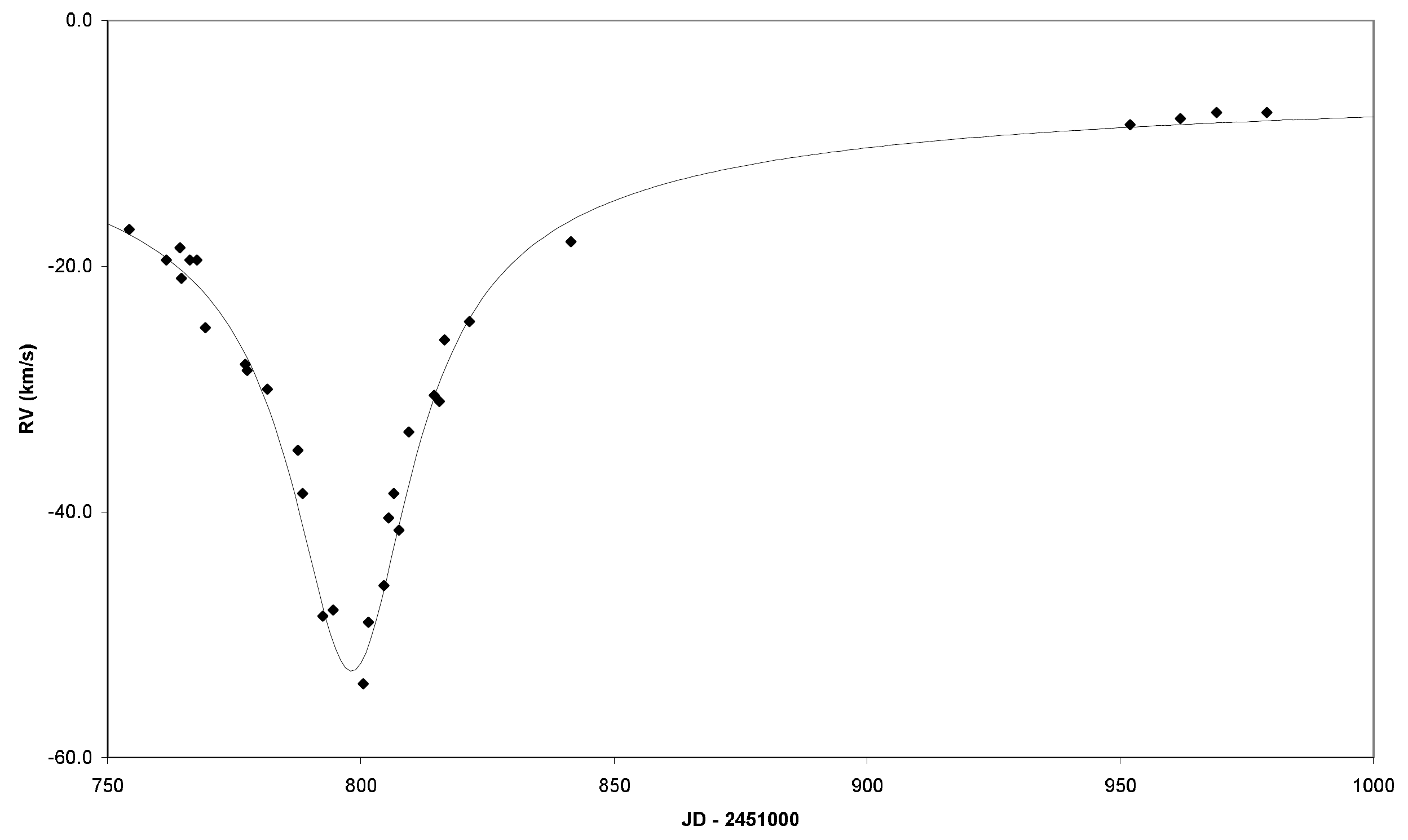}
\caption{The RV data and the RV curve for $\delta$ Sco calculated using
our revised elements (see Table \protect\ref{table:elements}). The
points are taken from Table 3 in \protect\citet{miro01} and we have
assumed that the differences between HJDs and JDs are
negligible.\label{fig:rv}}
\end{center}
\end{figure*}

All the interferometric measures ($\rho$, $\theta$ values) listed in
\citet{hartkopf04} were used as well as the measure reported here.
Where necessary the phases were adjusted by adding $180^\circ$ to make
them consistent with that found by \citet{bedding93} with
phase-closure. The Besselian epochs in the Fourth Catalog were also
converted to Julian epochs (this made a small but significant
improvement to the fit). A total of 36 interferometric measures were
used. The 30 RV values were taken from Table 3 in \citet{miro01}. The
interferometric data span the epoch J1973.2 to J1999.4; the
spectroscopic data are mostly concentrated around the epoch of
periastron, J2000.7, with four points measured near the epoch J2001.1.
The initial guesses for the orbital elements were taken from
\cite{miro01}; the initial guesses for the amplitude and the
$\gamma$-velocity were  $25$\,km$\cdot$s$^{-1}$ and
$-6$\,km$\cdot$s$^{-1}$, respectively. The results of our analysis are
shown in the last column of Table~\ref{table:elements}. This Table also
lists the elements previously obtained by \cite{hartkopf96} and
\cite{miro01}. The interferometric orbit is plotted in
Fig.~\ref{fig:orbit} along with the data and the ``observed minus
calculated'' (O-C) vectors. Fig.~\ref{fig:rv} shows the RV data and the
best fitting RV curve.

The most significant differences between our orbital fit and the
previous ones are in the semimajor axis and the period.  We found
$a^{\prime\prime} = 98.3\pm1.2$\,mas and $P=10.74\pm0.02$ yr; these
should be compared to the values of $a^{\prime\prime} = 107\pm7$\,mas
and $P=10.58\pm0.08$ yr found by \citet{hartkopf96} and assumed by
\citet{miro01}. The other orbital elements are in excellent agreement
with those found by \citet{miro01}, who inferred a value for $i$ from
the H$\alpha$ profile.

The most uncertain quantity is the mass function. The large uncertainty
in $f(m)$ is almost entirely due to $(\sin i)^{-3} = 4.3\pm1.9$.

\section{The masses of $\delta$ Sco A \& B}\label{sec:masses}

The magnitude difference between the two components is $\Delta m \sim 2$,
suggesting that the secondary is also an early type star. In this section we
examine this in more detail and estimate the masses of the two components.

Assuming that the two stars have the same age and chemical composition,
the luminosity and mass are related by the empirical mass-luminosity
relation \citep*{griffiths88}.
\begin{equation}
 \log_{10}(L_B/L_A) = (3.51\pm0.14)\log_{10}(M_B/M_A)
\end{equation}

We have previously used this method to determine the masses of the two massive
components of the $\lambda$ Sco system \citep{tango06}.  In that case there was
strong circumstantial evidence that the stars were formed at the same time: it
is a triple system and the orbits of the secondary and tertiary lie in the same
orbital plane. For $\delta$ Sco there is no additional information to support
the hypothesis that the two stars were formed together; indeed, the fact that
the orbit is so eccentric may be evidence that the system may have suffered
some kind of disruption in the past.

Because of the development of the emission region around the primary we also
need to treat the brightness ratio with caution. Although \citet{galaz06}
reported H$\beta$ emission prior to 1999 there is no indication of photometric
brightening and we believe that the emission region was optically thin in the
continuum at this time. This is supported by the data in
Table~\ref{table:magdiff}: there are no significant differences between the
$\Delta m$ measured at the three epochs 1971, 1991 and 1999.

The brightness ratio was measured at 442\,nm and we assume that $m_{442}$ is
equal to the B magnitude. To estimate the luminosity ratio we must know both
the B-V and bolometric corrections (BC) for the two stars, but the spectral
class of the secondary is unknown. We initially assumed that both stars had the
same  B-V and BC and used the mass-luminosity relation to make a preliminary
estimate of the masses. We then used the ZAMS model grid of \citet{tout96}
assuming solar metallicity to determine the effective temperatures from the
masses.    The B-V and bolometric corrections were then found by interpolating the
tables in \citet{flower96}. The luminosity ratio was found from the
experimentally determined brightness ratio $\beta_{442}$ using:
\begin{eqnarray}
    -2.5\log_{10}(L_B/L_A) &=& -2.5\log_{10}(\beta_{442})\nonumber\\
    &&- (B-V)_B + (B-V)_A \nonumber\\
    && + BC_B - BC_A
\end{eqnarray}
This was used to estimate the next approximation for the masses and the
procedure was repeated until the solution converged.  It should be noted that
the interpolation formulas given by \citet{tout96} made this quite
straightforward. 

Four iterations were sufficient and we estimate the mass ratio $q$ for
the two stars to be
\begin{equation}
    q = \frac{M_A}{M_B} = 1.957\pm0.011
\end{equation}
The individual masses can be determined from $f(m)$ and $q$:
\begin{eqnarray}
    M_A &=& 15 \pm 7 M_\odot\\
    M_B &=& 8.0 \pm 3.6  M_\odot
\end{eqnarray}
 
It should be noted that the choice of a ZAMS model grid may introduce systematic
errors as the $\delta$ Sco system may be somewhat evolved. However, the uncertainty in the masses of the stars is dominated by the uncertainty in the mass function  and consequently more detailed modelling is not justified at present.

The colour indices for the A and B components  found by the procedure
outlined above are B-V = $-0.29$ and $-0.25$, respectively, suggesting
that the spectral type of the secondary is approximately B2
\citep{LB82}. This is consistent with the observation by
\cite{carciofi06} that the spectroscopy suggests that the primary and
secondary have similar effective temperatures.

Given the masses, the dynamical parallax can be found from
\begin{equation}
    \varpi_d = \frac{a^{\prime\prime}}{P^{2/3}(M_A+M_B)^{1/3}} = 7.03\pm0.15\, {\rm mas}
\end{equation}

\section{Discussion}

The revised HIPPARCOS parallax $\varpi = 6.65\pm0.89$\,mas
\citep{hipparcos07} is in good agreement with our dynamical parallax.
Our parallax allows us to determine the physical size of the $\delta$
Sco disk;  we estimate its diameter to have been $\gtrsim
0.28\pm0.14$\,AU in 2000/2001 and $\gtrsim 0.57$\,AU in 2005/2006. The
physical separation of the two stars at periastron is $0.84$\,AU.

As noted previously, \citet{otero01} suggested that $\delta$ Sco might
be a $\gamma$ Cas type variable star. These are Be stars that exhibit
eruptive irregular variability, with light amplitudes reaching up to
1.5 in V \citep{samus04}.

The bright B0.5e star $\gamma$ Cas is also a binary system that is
slightly further away than $\delta$ Sco (its revised HIPPARCOS parallax
is $5.89\pm12$\,mas). It is well known for its unique and complex X-ray
emission that consists of several distinct components. The evidence
strongly suggests that the X-ray emission occurs at the Be star itself
or its disk \citep{smith04}.  The orbit was determined by \citet{harmanec00} to be moderately eccentric ($e=0.260$) whilst \citet{miro02} found the orbit to be circular. Although this discrepancy has yet to be resolved it is nevertheless clear that the binary characteristics of $\gamma$
Cas, however, are quite different to $\delta$ Sco; in particular its orbital period is approximately 200 days.   \citet{harmanec00} reported the mass of the $\gamma$ Cas companion to be $\sim
1M_\odot$ \citep{harmanec00}; according to \citet{smith04} the
secondary is most likely a dwarf and is not a white dwarf or neutron
star.   In comparison, $\delta$ Sco has a highly eccentric orbit, a period of 10.7 years and its companion  is a relatively massive B2 star.

X-ray emission typical of B stars had been detected from $\delta$ Sco
prior to the periastron events of 2000 \citep*{grillo92,berghofer96},
but we are not aware of any more recent X-ray observations. If $\delta$
Sco is analogous to $\gamma$ Cas we would now expect its X-ray spectrum
to exhibit some of the unusual features seen in the X-ray behaviour of
$\gamma$ Cas. Further work, particularly X-ray observations of this
intriguing system, are needed to clarify the nature of $\delta$ Sco.

\section*{Acknowledgments}

The authors gratefully thank the Australian Research Council and the University
of Sydney for supporting this work.

APJ and JRN acknowledge the support provided by a University of Sydney
Postgraduate Award. APJ also acknowledges the support provided by a Denison
Postgraduate Award.

We would like to thank the referee for his extensive and extremely
helpful criticisms of the original manuscript.

This research has made use of the SIMBAD database, operated at CDS, Strasbourg,
France.

%\bibliographystyle{apj}
%\bibliography{DeltaSco_final}

\label{lastpage}
\end{document}